\documentclass[a4paper,
               ]{jacow}
\usepackage{pdfpages,multirow,ragged2e} %
\makeatletter%
	\ifboolexpr{bool{xetex}}
	 {\renewcommand{\Gin@extensions}{.pdf,%
	                    .png,.jpg,.bmp,.pict,.tif,.psd,.mac,.sga,.tga,.gif,%
	                    .eps,.ps,%
	                    }}{}
\makeatother

\ifboolexpr{bool{xetex} or bool{luatex}} 
 {}                                      
 {\usepackage[utf8]{inputenc}}           

\usepackage[USenglish]{babel}

\ifboolexpr{bool{jacowbiblatex}}%
 {%
  \addbibresource{jacow-test.bib}
  \addbibresource{biblatex-examples.bib}
 }{}
\begin{document}

\title{Simulation of the dynamics of gas mixtures during plasma processing in the C75 Cavity
\thanks{This work is supported by SC Nuclear Physics Program through DOE SC Lab funding announcement Lab-20-2310 and by the U.S. Department of Energy, Office of Science, Office of Nuclear Physics under contract DE-AC05-06OR23177. }}

\author{N. K. Raut\thanks{raut@jlab.org}, T. Ganey, P. Dhakal, and T. Powers \\
 Thomas Jefferson National Laboratory, Newport News, VA 23606, USA 
}
	
\maketitle

\begin{abstract}
 Plasma processing using a mixture of noble gas and oxygen is a technique that is currently being used to reduce field emission and multipacting in accelerating cavities. Plasma is created inside the cavity when the gas mixture is exposed to an electromagnetic field that is generated by applying RF power through the fundamental power or higher-order mode couplers. Oxygen ions and atomic oxygen are created in the plasma which breaks down the hydrocarbons on the surface of the cavity and the residuals from this process are removed as part of the process gas flow. Removal of hydrocarbons from the surface increases the work function and reduces the secondary emission coefficient \cite{Tyagi}. This work describes the initial results of plasma simulation, which provides insight into the ignition process, distribution of different species, and interactions of free oxygen and oxygen ions with the cavity surfaces. The simulations have been done with an Ar/$O_2$ plasma using COMSOL® multiphysics. These simulations help in understanding the dynamics and control of plasma inside the cavity and the exploration of different gas mixtures. 
\end{abstract}

\section{Introduction}
Field emission in superconducting radio-frequency (SRF) cavities leads to thermal instability and is one of the prime factors in limiting the performance of accelerating cavities \cite{Padamsee}. Hydrocarbons (C$_x$H$_y$) build-up on the surface of the cavity enhance multipactors and field emission \cite{Bianca}. Particulate contamination is the major cause of field emission. Plasma helps to break down organic bonds (C$=$C, C$-$C, C$-$O, C$-$H) from the contamination \cite{Kim} and increases the work function ($\phi$) and secondary emission yield ($<SEY>$) of the niobium \cite{Doleans}. Recently, promising improvement on the onset of field emission and increase in usable accelerating gradient on SRF cavities \cite{Bianca, Powers, Martinello}.

\begin{figure}[htb]
   \centering
   \includegraphics*[width=.88\columnwidth]{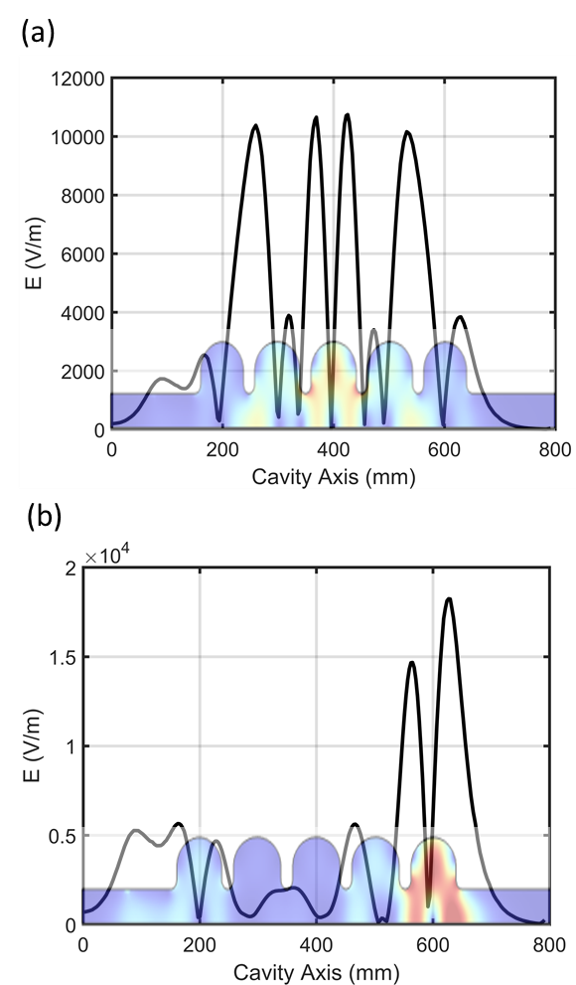}
   \caption{Electric field profile of two TE211 modes of interest on the axis of the C75 cavity, (a) 2656 MHz and (b) 2724 MHz. }
   \label{Fig1}
\end{figure}

In the plasma processing of the cavity, the reactive ions and species such as O$^-$, O$^+$, O, O$_2^+$ play an essential role to crack the hydrocarbons from the surface of the cavity forming residual byproducts such as CO, H$_2$O, CO$_2$, and etc. In the experimental settings, it is challenging to get the information about plasma and other species' growth in a fraction of a second as well as the interaction between the species with the cavity's surface.  Furthermore, the generation of plasma  with the optimum proportion of the gases mixture (n$\%$ of inert gas $\&$ (100-n)$\%$ of oxygen) requires careful control of the gas mixture and plasma dynamics. Experimentally, optical camera attached on cavity opening is used to observe the plasma ignition and its evolution. Simulation of plasma ignition, dynamics with respect to the partial pressure of gas and rf power could be a useful tool to design the experimental setup in complex cryomodules where visual observations are not available.


\section{Computational Model}
 In this study, we have chosen two quadrupole mode resonating at 2656 MHz and 2724 MHz of the C75 cavity \cite{Marhauser} to ignite plasma on the center and end cells of the cavity. COMSOL Multiphysics has been implemented to study the interaction between the cavity's modes and the gaseous mixture. Here, we report simulation results on electron number density (N$_e$), temperature (T$_e$), S-parameters of the cavity ($S11 \& S21$), and dynamics of the oxygen species on the axis \& surface of the cavity. 

Figure~\ref{Fig1} shows the electric field distribution of two TE211  modes on the axis of the C75 cavity used for the plasma simulation. A detailed discussion about the C75 cavity is done in Ref. \cite{Marhauser}. Figure~\ref{Fig1} (a) shows the 2656 MHz mode profile with most of its field around the center cells of the cavity.  The second mode is the 2724 MHz mode (see Fig.~\ref{Fig1} (b)), which has the highest field on the end cell. In the simulation, the electric field of the cavity is excited via a coaxial input port. For the sake of simulation time, 2D axis-symmetric part of the cavity has been used. 

Oxygen plasma is highly reactive due to high concentrations of active particles and electronically excited metastable states. In this simulation, a gas mixture of 94 \% Ar and 6 \%  of O$_2$ is set within the cavity domain. When there is an interaction between the electromagnetic field and gas molecules, electrons absorb energy from the electric field and lose it to the gas molecules. During this repetitive collision process between the excited electrons and the gas molecules, highly reactive ions and metastable species along with the electrons and neutral atoms or molecules are produced. This phase of gas inside the cavity is called plasma ignition. The E-field profiles that are shown in Fig.~\ref{Fig1} (a) \& (b) ignite the plasma, respectively, at the center and end cell of the cavity. 

The COMSOL Multiphysics for plasma simulation solves drift-diffusion equations to calculate the transport properties of the electron and non-electronic species.  The plasma chemistry is described and represented by the different reactions and divided into Tables 1 and 2, representing the electron impact reactions of Ar and O$_2$. We have used 5 such reactions of Ar and 35 of O$_2$. These reactions are taken from the database system LxCat \cite{lxcat}. The reactions are mainly elastic, attachment, excitation, and ionization. The attachment, excitation, and ionization reactions are used in plasma processing to react with the hydrocarbons on the cavity surface. 

\begin{table} [hbt]
\centering
   \caption{Argon Reactions \cite{lxcat}}
   \begin{tabular}{lccc}
       \toprule
\textbf{\#} & \textbf{Formula}  & \textbf{Type} & \textbf{$\Delta \epsilon$} (eV)\\
       \midrule
1	& {$e+Ar=>e+Ar$}	& Elas.$^1$	& 0  \\ 
2	& {$e+Ar=>e+Ars$}	& Exct.$^2$	& 11.5 \\ 
3	& {$e+Ars=>e+Ar$}	& Exct.$^2$	& -11.5 \\ 
4	& {$e+Ar=>2e+Ar^+$}	& Ion.$^3$	& 15.8 \\ 
5	& {$e+Ars=>2e+Ar^+$}	& Ion.$^3$	& 4.427 \\ 
       \bottomrule  
   \end{tabular}   \\
      $^1$Elastic, $^2$Excitation, $^3$Ionization  
   \label{table1}
 \end{table} 
 
 In the elastic reactions, there is an energy exchange between the electron and Ar or O$_2$ molecules. However, no new species are created in this process. In the attachment reactions, electrons can be taken by the species to form such reactive species. In addition, reactions like ionization can also produce reactive species. In these reactions, an interacting electron knocks out an electron from participating species by releasing energy. Ionic species like Ar$^+$, O$_2^+$, and O$^+$ are created in the processes. Moreover, the electrons provide their energy to the ground state Ar and O$_2$ resulting in new excited species such as Ars, O$_2a1d$, O$_2b1s$,  O$_2 (45)$, O1d, O1s, and  O.  
 
\begin{table} 
\centering
   \caption{Oxygen Electron Impact Reactions \cite{lxcat}}
   \begin{tabular}{lccc}
       \toprule
\textbf{\#} & \textbf{Formula}  & \textbf{Types} & \textbf{$\Delta \epsilon$} (eV)\\
       \midrule
1    & {$e+O_2=>e+O_2$}   & Elas.$^1$  & 0  \\ 
2    & {$e+O_2=>O+O^-$}   & Att.$^4$   & $-$  \\ 
3    & {$e+O_2=>e+O_2$}   & Exct.$^2$  & $0.02$  \\ 
4    & {$e+O_2=>e+O_2$}   & Exct.$^2$   & $0.19$  \\ 
5    & {$e+O_2=>e+O_2$}   & Exct.$^2$   & $0.38$  \\ 
6	& {$e+O_2=>e+O_2$}	& Exct.$^2$	& 0.57 \\ 
7	& {$e+O_2=>e+O_2$}	& Exct.$^2$	& 0.75 \\ 
8	& {$e+O_2=>e+O_2a1d$}	& Exct.$^2$	& 0.977 \\ 
9	& {$e+O_2a1d=>e+O_2$}	& Exct.$^2$	& -0.977 \\ 
10	& {$e+O_2=>e+O_2b1s$}	& Exct.$^2$	& 1.627 \\ 
11	& {$e+O_2b1s=>e+O_2$}	& Exct.$^2$	& -1.627 \\ 
12	& {$e+O_2=>e+O_2(45)$}	& Exct.$^2$	& 4.5 \\ 
13	& {$e+O_2(45)=>e+O_2$}	& Exct.$^2$	& -4.5 \\ 
14	& {$e+O_2=>e+O+O$}	& Exct.$^2$	& 6.0 \\ 
15	& {$e+O_2=>e+O+O1d$}	& Exct.$^2$	& 8.4 \\ 
16	& {$e+O_2=>e+O+O1s$}	& Exct.$^2$	& 9.97 \\ 
17	& {$e+O_2=>2e+O_2^+$}	& Ion.$^3$	& 12.06 \\ 
18	& {$e+O_2a1d=>e+O_2a1d$}	& Elas.$^1$ & 0 \\ 
19	& {$e+O_2a1d=>e+O+O$}	& Exct.$^2$	& 5.02 \\ 
20	& {$e+O_2a1d=>2e+O_2^+$}	& Ion.$^3$	& 11.09 \\ 
21	& {$e+O_2b1s=>e+O_2b1s$}	& Elas.$^1$ & 0 \\ 
22	& {$e+O_2b1s=>e+O+O$}	& Exct.$^2$	& 4.38 \\ 
23	& {$e+O_2b1s=>2e+O_2^+$}	& Ion.$^3$	& 10.39 \\ 
24	& {$e+O_2(45)=>e+O+O$}	& Exct.$^2$	& 1.5 \\ 
25	& {$e+O_2(45)=>2e+O_2^+$}	& Ion.$^3$	& 7.58 \\ 
26	& {$e+O=>e+O$}	& Elas.$^1$	& 0  \\ 
27	& {$e+O=>e+O1d$}	& Exct.$^2$	& 1.968  \\ 
28	& {$e+O1d=>e+O$}	& Exct.$^2$	& -1.968 \\ 
29	& {$e+O=>e+O1s$}	& Exct.$^2$	& 4.192 \\ 
30	& {$e+O1s=>e+O$}	& Exct.$^2$	& -4.192 \\ 
31	& {$e+O=>2e+O^+$}	& Ion.$^3$	& 13.192 \\ 
32	& {$e+O1d=>e+O1s$}	& Exct.$^2$	& 2.224 \\ 
33	& {$e+O1d=>2e+O^+$}	& Ion.$^3$	& 11.224  \\ 
34	& {$e+O1s=>2e+O^+$}	& Ion.$^3$	& 9 \\ 
35	& {$e+O_2+O_2=>O_2+O_2^-$}	& Att.$^4$	& 0 \\ 
 \bottomrule
   \end{tabular}   
   $^1$Elastic, $^2$Excitation, $^3$Ionization, $^4$Attachment
   \label{table1}
 \end{table}

\begin{figure}[t!]
\centering
\qquad
  \includegraphics[width=0.75\linewidth]{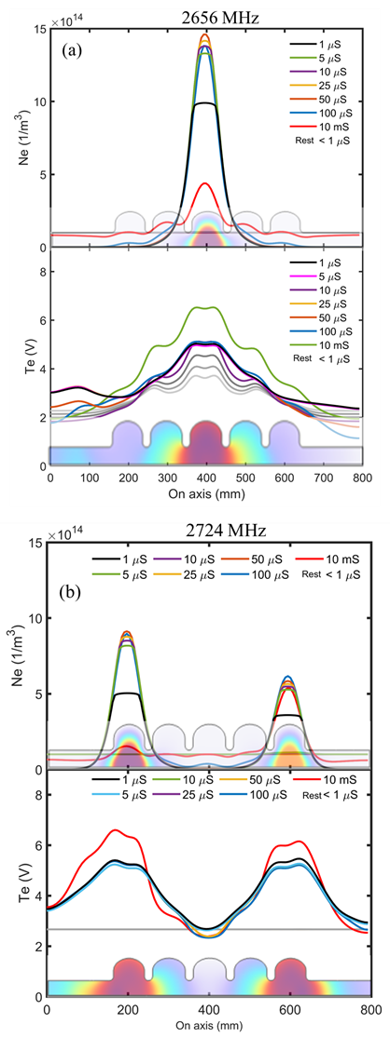}
 \caption{Electron number density ($N_e$) as a function of time for (a) 2656 MHz, and (b) 2724 MHz mode of the cavity. For visualization purpose the 2D-axis symmetric cavity is inserted in both plots.}
    \label{NeTe}
\end{figure}

\section{RESULTS AND DISCUSSION}
\subsection{Electron Number Density and Temperature}
 Two of the important parameters of the plasma simulation are growth in the electron number density (N$_e$ ) and heating of the gas molecules (T$_e$). An increase in the interaction between the gas molecules and the electric field of the cavity results in a growth of the number of free electrons and a rise in temperature. Figure~\ref{NeTe} shows the change in the N$_e$ and T$_e$ as functions of time. Here, the data are extracted on the axis of the cavity. To increase the resolution of the results only cell-to-cell calculations are included in the plots. 
 
 \begin{figure}[!tbh]
    \centering
    \includegraphics[width=\linewidth]{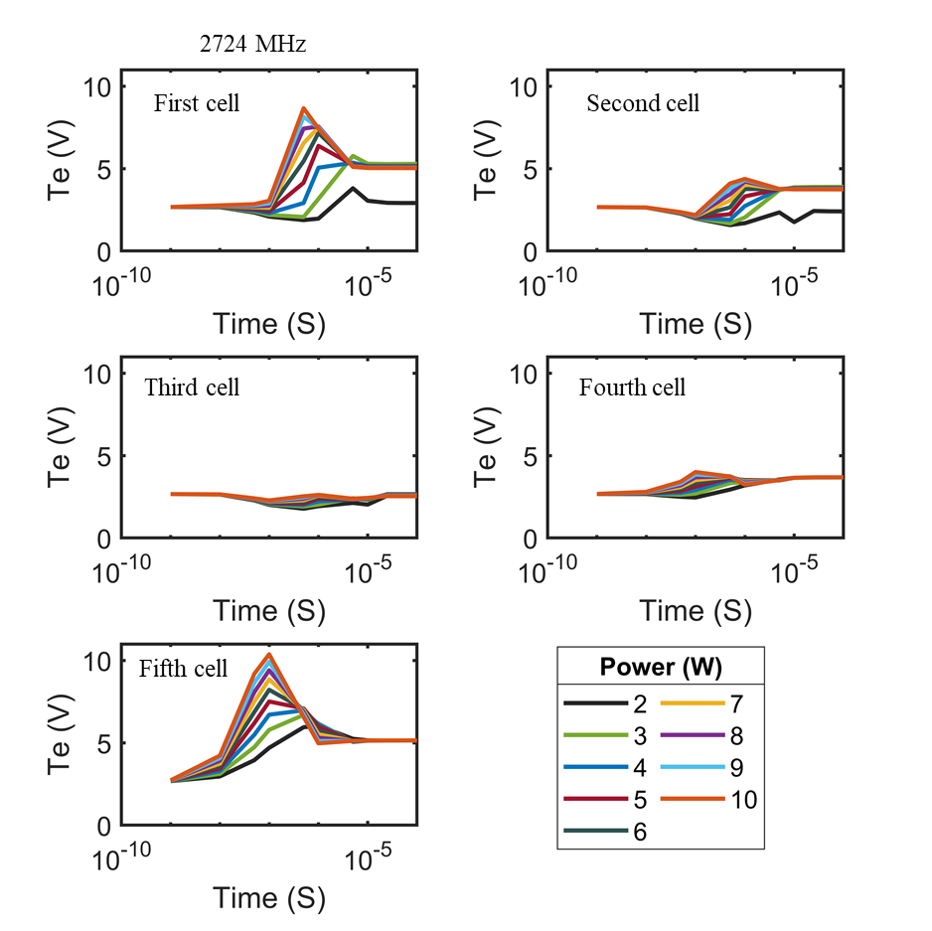}
    
    \caption{Electron temperature at center points of five cells of the cavity for 2724 MHz for powers 2 - 10 W. The red dots show the locations of the data point inside the cavity.}
    \label{Te1}
\end{figure}


Figure~\ref{NeTe} (a) and (b) displays (N$_e$, T$_e$), for 2656 MHz and 2724 MHz mode of the cavity, respectively. In both cases, a significant increase in N$_e$ and T$_e$ is observed. In case of 2656 MHz, N$_e$ and T$_e$ increased from 10$^{14}$ m$^{-3}$ and 2.7 V to the maximum value of 1.4$\times$10$^{15}$ m$^{-3}$ and 6.6 V at the center cell of the cavity. However, at 2724 MHz, there is a rise in (N$_e$, T$_e$) not only at the end cell but also in the first cell of the cavity (see Figure ~\ref{NeTe} (b)). In this mode, for a time greater than 1 microsecond, the growth of N$_e$ in the first cell is higher than that of the last cell. Similar built-up behavior in T$_e$ is also observed. 

To understand the shifting behavior of T$_e$ from the end cell to the first cell at 2724 MHz, we have done simulations at different input powers from 2 - 10 W as shown in Fig. ~\ref{Te1}. The heating of the gas molecules on the end and the first cell of the cavity and their neighboring cells was observed. However, there is negligible change in T$_e$ on the center cell. 
 
\subsection{S-parameters}
For plasma to propagate inside the cavity, the plasma density ($n_p$) should be less than the critical density ($n_c$) due to the Debye shielding \cite{Brown}. The value of the critical density is determined by the angular frequency of the cavity ($\omega$), electronic charge ($e$), and mass ($m_e$) and is calculated as;
\begin{equation}\label{eq:units}
   n_c=\frac{\epsilon_0 m_e^2 \omega^2}{e^2}
\end{equation}
\begin{figure}[htb]
   \centering
   \includegraphics*[width=.85\columnwidth]{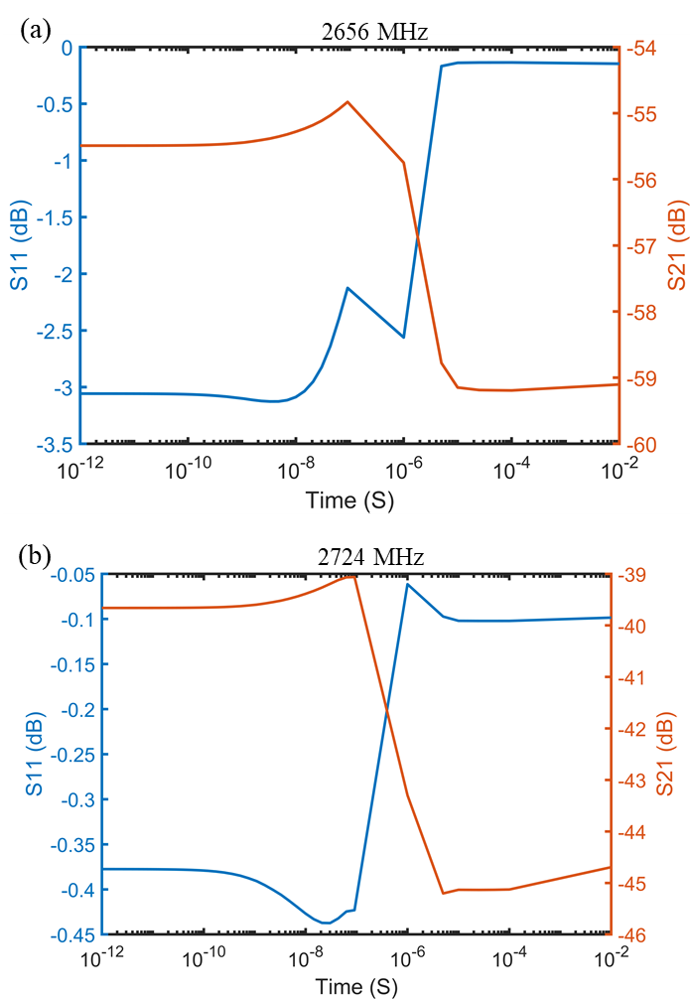}
   \caption{Change in the S-parameters of the cavity as a function of time for (a) 2656 MHz and (b) 2724 MHz.}
    \label{Spara}
\end{figure}
In the experimental setting, one of the important parameters to track during plasma ignition is the change in the S-parameters (S11 and S21) of the cavity \cite{Ahmed}. Plasma development changes the dielectric constant of the medium as \cite{Brown}:
\begin{equation}\label{eq:units}
   \epsilon_r=1-\frac{\omega_p^2}{\omega^2(1-i\frac{\nu}{\omega})}
\end{equation}
where $\omega_p$ and $\omega$ are the frequency of the plasma and the electromagnetic field, respectively and  $\nu$ is the wavenumber. An increase in the electron number density (N$_e$) was observed as the plasma starts to form, which increases plasma conductivity and hence the frequency. The increase in the plasma frequency in turn changes the dielectric constant of the cavity resulting in the increase of the reflection of the electromagnetic waves inside the cavity. 

Figure~\ref{Spara} shows the cavity's S-parameters (S11 and S21) as a function of time for both modes used in this simulation. In both modes decrease in S21 by 5 dB during the plasma ignition is observed. During the ignition, as expected, there was an increase in the S11 parameters of the cavity. A similar trend of change in S-parameters of the C100 cavity is reported by \cite{Powers}.  The C100 cavity is 7 cells cavity and the change in S21 was 10-20 dB. 

\subsection{Species Dynamics}
In plasma processing, free oxygen radicals and ions play a crucial role in the breakdown of hydrocarbons from the surface of the cavity to molecules such as CO, CO$_2$, and H$_2$O because of their highly reactive nature. An oxygen molecule reacts with a hydrocarbon as O$_2$ + C$_x$H$_y$$\rightarrow$ $CO_2$ + $CO$+$H_2O$. 
\begin{figure}[!htb]
   \centering
   \includegraphics*[width=.9\columnwidth]{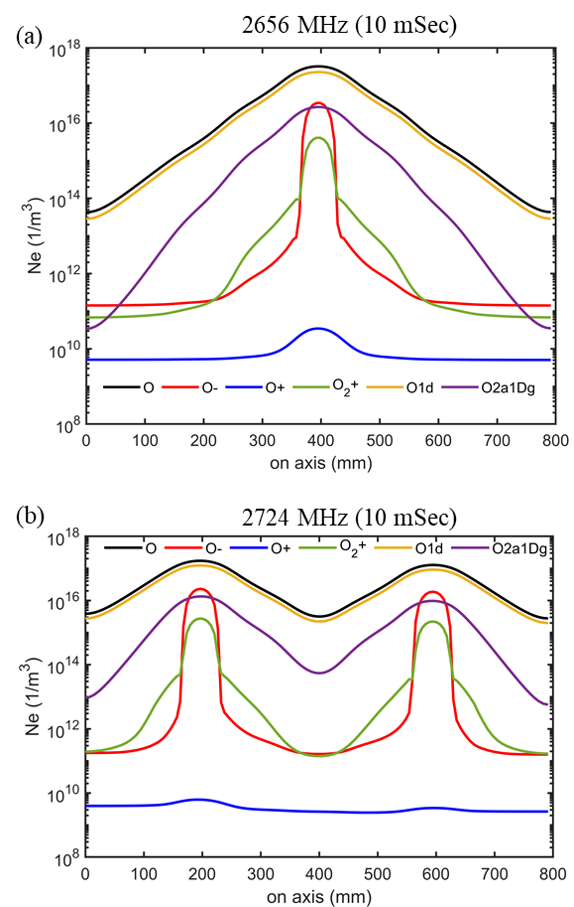}
   \caption{Different species of oxygen (molecules, ions, \&  metastable states) on the axis of the cavity for (a) 2556 MHz and (b) 2772 MHz. }
    \label{O-O+}
\end{figure}

Figure~\ref{O-O+} shows a change in O molecule, O$^-$, O$^+$, \& O$_2^+$ ions and (O1d, O2a1Dg) metastable states of the oxygen at 10 ms of time for the two modes of interest. In both modes, we have observed an increase in all of the species mentioned above.  Among all, there is a substantial increase in O$^-$ ions. Interestingly, in 2724 MHz mode, we have seen simultaneous growth of the species at the end and first cell of the cavity. 

\begin{figure}[htb]
   \centering
   \includegraphics*[width=1\columnwidth]{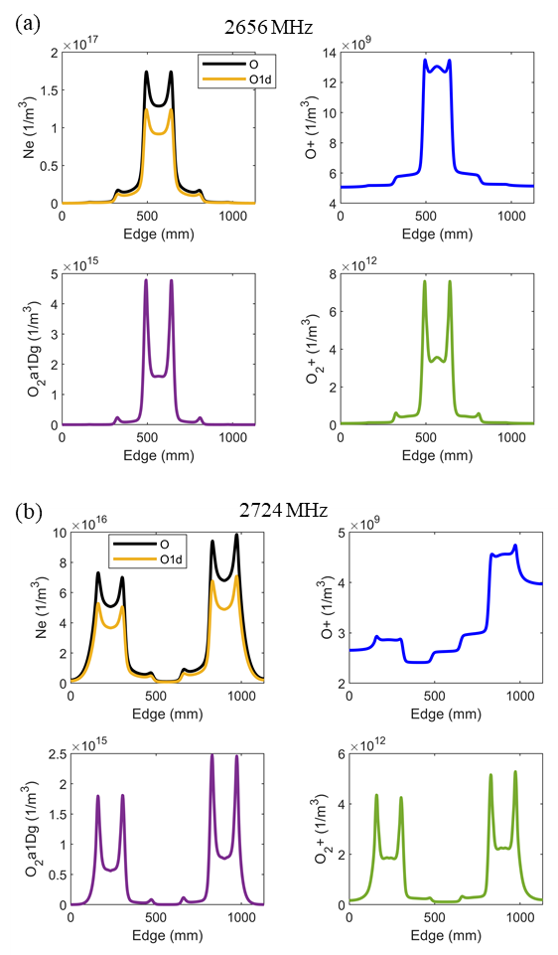}
   \caption{Oxygen species O, O1d, O$^+$, O$_2a1Dg$, and O$_2^+$ along the edge of the cavity for two modes used in this simulation: (a) 26565 MHz and (b) 2724 MHz. }
    \label{edge_species}
\end{figure}

Calculations of species dynamics along the inner surface of the cavity are done and shown in Fig.~\ref{edge_species}. As expected, for the center cell plasma ignition, there was the growth of the (O, O1d, O$^+$, O$_2a1Dg$, and O$_2^+$) along the circumference of the center cell of the cavity (see Fig.~\ref{edge_species} (a)). Here, two peaks of species at two ends of the same cell of the cavity is due to the feature of the quadrupole mode. Similar to the shifting feature of species seen on the axis of the cavity, for 2724 MHz mode, the species was created on the circumferences of both the end and first cell of the cavity. This feature suggests that plasma processing at the end cell is more effective to clean cavities than that of center cell. 

\section{CONCLUSION}
COMSOL multiphysics has been used to study plasma ignition and species growth in the C75 cavity. A mixture of 94:6 Ar and $O_2$ is set inside the cavity domain and two TE211 modes are used to ignite plasma on the center and end cell of the cavity, respectively. A significant increase in the free electron number density and their heating shows the creation of the plasma.  Moreover, we find a 5 dB decrease in the S21 parameter of the cavity during the plasma ignition. The study of oxygen molecules, ions, and their metastable states on the axis and inner surface of the cavity suggests the end cell plasma ignition could be more effective than the center cell plasma ignition because of the simultaneous growth of species in both the first and end cell of the cavity. Our next step is to study plasma ignition on the second and fourth cell of the cavity and remodeling the EM fields due to the change in dielectric constant then repeat the plasma simulation with an initial condition of the final condition of the previous simulation.

This work could guide to study plasma ignition and control in any shape and size of accelerating cavities. It could also be a platform to understand species growth and their dynamics during plasma ignition. 

\section{ACKNOWLEDGEMENTS}
We would like to acknowledge Jefferson Lab SRF S\&T department for support. 

%
%
\ifboolexpr{bool{jacowbiblatex}}%
	{\printbibliography}%
	{%
	
	
%
%

}
\end{document}